\documentclass[review]{elsarticle}

\usepackage{lineno,hyperref}

\pdfoutput=1
 
\usepackage[T1]{fontenc}
\usepackage{ae,aecompl}
\usepackage{lmodern}
\usepackage[utf8]{inputenc} 
\usepackage{graphicx}
\usepackage{dcolumn}
\usepackage{bm}
\usepackage{hyperref}
\usepackage{subfigure}
\usepackage{longtable}
\usepackage{lipsum}
\usepackage{color}
\usepackage{comment}
\usepackage[top=1in, bottom=1in, left=1in, right=1in]{geometry}

\usepackage{amsmath}
\usepackage{amssymb}
\usepackage{dsfont}

\newcommand{\Nsp}{{N_\mathrm{sp}}}
\newcommand{\Sww}{{\boldsymbol{S}_{\boldsymbol{w}\boldsymbol{w}}}}
\newcommand{\Swv}{{\boldsymbol{S}_{\boldsymbol{w}v}}}
\newcommand{\Swweq}{ {\boldsymbol{S}^\mathrm{eq}_{\boldsymbol{w}\boldsymbol{w}} } }
\newcommand{\Swveq}{ {\boldsymbol{S}^\mathrm{eq}_{\boldsymbol{w}v} } }
\newcommand{\SwweqO}{ {\boldsymbol{S}^{\mathrm{eq,n}}_{\boldsymbol{w}\boldsymbol{w}} } }

\newcommand{\SwwP}{{\boldsymbol{S}^\prime_{\boldsymbol{w}\boldsymbol{w}}}}
\newcommand{\SwvP}{{\boldsymbol{S}^\prime_{\boldsymbol{w}v}}}

\global\long\def\V#1{{\boldsymbol{#1}}}
\global\long\def\M#1{\boldsymbol{#1}}

\global\long\def\d#1{\delta#1}

\global\long\def\grad{\M{\nabla}}

\newcommand{\modified}[1]{#1}
\newcommand{\deleted}[1]{}

\newcommand{\Donev}[1]{}   
\newcommand{\Garcia}[1]{}  
\newcommand{\JPP}[1]{}  

\newcommand{\sidenote}[1]{{\color{blue}#1}}

\begin{document}
\begin{frontmatter}

\title{Fluctuating Hydrodynamics and Debye-H\"uckel-Onsager Theory for Electrolytes}

\author{Aleksandar Donev}
\ead{donev@courant.nyu.edu}
\address{Courant Institute of Mathematical Sciences, New York University, New York, NY, 10003}

\author{Alejandro L. Garcia}
\address{Department of Physics and Astronomy, San Jose State University, San Jose, CA, 95192}

\author{Jean-Philippe P\'eraud, Andy Nonaka, John B. Bell}
\address{Center for Computational Science and Engineering, Lawrence Berkeley National Laboratory, Berkeley, CA, 94720}

\date{\today}

\begin{abstract}
We apply fluctuating hydrodynamics to strong electrolyte mixtures to compute the concentration corrections for chemical potential, diffusivity, and conductivity. We show these corrections to be in agreement with the limiting laws of Debye, H\"uckel, and Onsager. We compute explicit corrections for a symmetric ternary mixture and find that the co-ion Maxwell-Stefan diffusion coefficients can be negative, in agreement with experimental findings.   
\end{abstract}



\end{frontmatter}


\section{Introduction}

Due to the long-range nature of Coulomb forces between ions it is well-known that electrolyte solutions have unique properties that distinguish them from ordinary mixtures~\cite{dill2012molecular}.
Colligative properties, such as osmotic pressure, and transport
properties, such as mobility, have corrections that scale with the square root of concentration~\cite{robinson2012electrolyte,newman2012electrochemical,wright2007introduction,CouplingIonicDiffusion_Krishna}.  
This macroscopic effect has a mesoscopic origin, specifically, due to the competition of thermal and electrostatic energy at scales comparable to the Debye length.

The traditional derivation of the thermodynamic corrections is by way of solving the Poisson-Boltzmann equation. For example, an approximate solution gives the Debye-H\"uckel limiting law for the activity coefficient~\cite{robinson2012electrolyte,Debye1923}. The derivation of the transport properties, as developed by Onsager and co-workers~\cite{OnsagerFuoss1932,OnsagerKim1957,Electrolytes_OnsagerChen}, has a similar starting point but is much more complicated.
Here, we present an alternative approach using fluctuating hydrodynamics (FHD) \cite{FluctHydroNonEq_Book}.
\modified{
This paper generalizes our previous derivation for binary electrolytes~\cite{PnasElectrolyte} to arbitrary solute mixtures, and, as an illustrative example, calculates transport properties for a ternary electrolyte.
It should be noted that our FHD approach extends closely-related density functional theory calculations of relaxation corrections to the conductivity~\cite{DDFT_Electrolytes_Dean} of binary electrolytes to account for advection, which enables us to also compute the electrophoretic corrections~\cite{robinson2012electrolyte}. 
}

First formulated by Landau and Lifshitz to predict light scattering spectra \cite{Landau:Fluid,berne_pecora}, more recently FHD has been applied to study various mesoscopic phenomena in fluid dynamics \cite{FluctHydroNonEq_Book,DiffusionJSTAT,FractalDiffusion_Microgravity,Croccolo2016}. The current popularity of fluctuating hydrodynamics is due, in part, to the availability of efficient and accurate numerical schemes for solving the FHD equations~\cite{LowMachExplicit,LowMachImplicit,LowMachMultispecies,FluctReactDiff,LowMachElectrolytes,FHD_IonicLiquids, StagerredFluctHydro, StaggeredFluct_Energy,AMR_ReactionDiffusion_Atzberger}.
While in this work we give analytical results for dilute electrolytes under simplifying assumptions, numerical techniques can in principle be used to compute the transport coefficients for moderately dilute solutions.

From the work of Onsager \emph{et al.}~\cite{OnsagerFuoss1932,OnsagerKim1957,Electrolytes_OnsagerChen} we can obtain the Fickian diffusion matrix for sufficiently dilute electrolytes. \modified{For neutral multispecies mixtures, especially non-dilute ones, using binary Maxwell-Stefan (MS) diffusion coefficients (inverse friction coefficients)~\cite{MaxwellStefan_Review,ElectrolytesMS_Review} is preferred
because generally they are positive and} depend weakly on concentration. \modified{In this paper we use our FHD formulation to calculate the effective macroscopic (renormalized) co-ion and counter-ion MS coefficients for binary and symmetric ternary electrolytes.}
For sufficiently dilute binary electrolytes both theory and experiments show that the counter-ion MS friction coefficient diverges as the inverse square root of the ionic strength~\cite{ElectrolytesMS_Review}. This strong concentration dependence implies that there are cross-diffusion terms that couple the electrodiffusion of the different species even in the absence of an (external) electric field. Such effects are not captured in the widely-used Poisson-Nernst-Planck (PNP) model of electrodiffusion, which is only accurate for very dilute solutions. Our results show that for a symmetric ternary mixture all MS friction coefficients diverge as the inverse square root of the ionic strength and the co-ion coefficient can be negative, in agreement with experimental findings.

Obtaining MS coefficients as a function of concentration is quite difficult for multicomponent electrolytes,
and in practice various empirical fits are used following the work of Newman and collaborators~\cite{newman2012electrochemical}, especially in modeling electrodes in Li-ion batteries ~\cite{TernaryElectrolyteMS_LiPF6,ElectrolyteMS_LiPF6,doyle1993modeling,smith2017multiphase}. These fits have sometimes been informed by tracer diffusion coefficients of ions in binary electrolyte solutions~\cite{ElectrolytesMS_Pinto}. Some authors have computed MS coefficients by Green-Kubo formulas using molecular dynamics~\cite{Electrolytes_MS_1,Electrolytes_MS_2}. Asymptotic expansions for dilute solutions can be computed using our FHD-based approach and thus ground empirical fits of the concentration dependence used in chemical engineering.

A complete model of transport in electrolyte mixtures must also account for advection and thus include the momentum conservation (velocity) equation. This is commonly not done for dilute solutions based on arguments that the velocities are negligible (small Peclet number), but these arguments have often been flawed. This is because there is a Lorentz force in the momentum equation that induces nontrivial velocities even in the electroneutral bulk~\cite{ElectroneutralAsymptotics_Yariv}. Here we demonstrate that the coupling between charge fluctuations and velocity fluctuations via the Lorentz force is responsible for the so-called electrophoretic correction to the diffusion coefficients; it is this correction that can make the co-ion MS coefficient negative. Even more unexpected couplings between mass and momentum transport have been uncovered in double layers, where charges are nonzero and applied electric fields can introduce pressure gradients that then drive nontrivial barodiffusion~\cite{BarodiffusionElectrolytes}.

This paper is organized as follows. Section~\ref{GeneralFhdSection} presents the FHD equations for a strong electrolyte; we show that the equilibrium solution leads to the Debye-H\"uckel limiting law for activity. The non-equilibrium solutions derived in Section~\ref{RenormTransportSection} yield the relaxation and electrophoretic corrections to conductivity and diffusion originally derived by Onsager~\cite{OnsagerFuoss1932,OnsagerKim1957}. An additional renormalization correction due to correlations of concentration and velocity fluctuations is also derived~\cite{DiffusionJSTAT}. Section~\ref{sec:ternary} highlights an interesting co-ion cross-diffusion effect found in a ternary mixture. Section~\ref{ConclusionSection} outlines FHD applications for electrolytes that go far beyond re-deriving classical results.

\section{Fluctuating hydrodynamics for electrolytes}
\label{GeneralFhdSection}

We consider an electrolyte solution with $\Nsp$ solute species and 
let $w_i(\V{r},t)$ be the mass fraction for species $i$ at position $\V{r}$ and time $t$. The charge of a molecule (ion) is $e V_i$, where $V_i$ is the valence and $e$ is the elementary charge; we also write it as $m_i z_i$ where $m_i$ is the molecule mass and $z_i$ is the specific charge. We denote the vectors $\V{w} = (w_1,\ldots,w_\Nsp)^T$ and similarly $\V{z} = (z_1,\ldots,z_\Nsp)^T$ \modified{, where $(\cdot)^T$ denotes transpose.}

The fluid mixture is assumed incompressible with constant mass density $\rho$, isothermal with temperature $T$, and, on average, locally electroneutral ($\sum_i z_i \langle w_i \rangle = 0$).

\subsection{Stochastic transport equations}

For dilute ionic solutions, the transport (conservation) equation for solute species $i$ is
\begin{equation}
\partial_t w_i = 
-\grad \cdot (\V{F}_i + \widetilde{\V{F}}_i),
\label{MassTransportEqn}
\end{equation}
where $\V{F}_i$ is the hydrodynamic (dissipative and advective) flux, and $\widetilde{\V{F}}_i$ is the stochastic flux.
The diffusive flux is given by the Nernst-Planck equation so the total hydrodynamic flux is
\begin{equation}
\V{F}_i = - D_i^0 \left( \grad w_i + \frac{e V_i w_i}{k_B T}  \grad \phi \right) + \V{v} w_i,
\label{MassFluxEqn}
\end{equation}
where $\V{v}$ is the fluid velocity, $D_i^0$ is the ``bare'' Fickian diffusion coefficient,\footnote{The term bare refers to the fact that $D_i^0$ will later be renormalized by the fluctuations to its macroscopic value $D_i$.} $\phi$ is the electric potential, and $k_B$ is Boltzmann's constant. The dielectric permittivity $\epsilon$ is taken as constant so $\phi$ is defined by the electrostatic equation
\begin{equation}
- \epsilon\nabla^2 \phi = q, 
\qquad\mathrm{where}\qquad
q = \rho e \sum_i \frac{w_i V_i}{m_i}
= \rho \sum_i w_i z_i
\label{PoissonEquation}
\end{equation}
is the charge density. 

The stochastic species flux is
\begin{equation}
\widetilde{\V{F}}_i = \sqrt{\frac{2 D_i^0 m_i w_i}{\rho}} ~\V{\mathcal{Z}}_i,
\label{MassNoiseEqn}
\end{equation}
where $\V{\mathcal{Z}}$ is a Gaussian white noise vector field with independent components that are uncorrelated in time and space. This flux has zero mean ($\langle \widetilde{\V{F}}_i \rangle = 0)$ and its variance satisfies the fluctuation-dissipation theorem~\cite{FluctHydroNonEq_Book}.

We will refer to (\ref{MassTransportEqn},\ref{MassFluxEqn},\ref{PoissonEquation},\ref{MassNoiseEqn}) as the (fluctuating) Poisson-Nernst-Planck (PNP) equations \cite{LowMachElectrolytes}. 
Note that summing \eqref{MassTransportEqn} over all species gives the continuity equation $\grad\cdot\V{v}=0$.

The equation for momentum transport is
\begin{equation}
\rho \partial_t \V{v} = - \rho \grad \cdot (\V{v} \V{v}^T) - \grad p + \mu \nabla^2 \V{v}  + q \V{E} + \sqrt{\mu k_B T}~ \grad \cdot (\M{\mathcal{V}} + \M{\mathcal{V}}^T),
\label{MomentumTransportEqn}
\end{equation}
where $p$ is pressure, $\mu=\nu \rho$ is the shear viscosity, and  $\V{E}=-\grad \phi$ is the electric field.
The last term in~\eqref{MomentumTransportEqn} is the divergence of the stochastic stress tensor, where $\M{\mathcal{V}}$ is a white noise tensor field.

\subsection{Structure factor}

The static structure factor $S_{fg}(\V{k})$ characterizes the cross-correlations between the fluctuations of two scalar quantities $f(\V{r})$ and $g(\V{r})$,
\begin{equation}
S_{fg}(\V{k})=\langle \delta\hat{f}(\V{k}) \delta\hat{g}(\V{k})^* \rangle    
\end{equation}
where $\hat{f}(\V{k})$ is the Fourier transform of $f(\V{r})$ and $(\cdot)^*$ denotes conjugate transpose. 
By Plancherel's theorem, 
\begin{equation}
\langle (\delta{f}) (\delta{g})^* \rangle
= \frac{1}{(2\pi)^3} \int d\V{k} ~ S_{fg}(\V{k}).
\label{PlancherelEquation}
\end{equation}
Here the quantities of interest are the fluctuations of the mass fractions $\delta w_i = w_i - \bar{w}_i$ from their average $\bar{w}_i = \langle w_i \rangle$, and the fluctuations of the fluid velocity\footnote{We take the average fluid velocity as zero so $\delta \V{v} = \V{v}$; the notation emphasizes that velocity is a fluctuating quantity.} $\delta \V{v}$. 
In the non-equilibrium situations considered here we are only interested in the velocity component in the direction of the applied thermodynamic force (e.g., external electric field). This is taken as the $x$-direction so only $v_x$ is retained in the structure factors.

A central quantity in our calculations is the $(\Nsp + 1) \times (\Nsp + 1)$ Hermitian matrix of structure factors
\begin{equation}
\V{S} = 
\left(\begin{array}{c |c}
\Sww & \Swv \\
\hline
{\Swv}^* & S_{vv}
\end{array}\right),
\end{equation}
where the matrix $\Sww = \langle (\delta\hat{\V{w}})( \delta\hat{\V{w}})^* \rangle$ and the vector $\Swv$ have elements
\begin{equation}
[\Sww]_{ij} = S_{w_i,w_j} = \langle (\d{\hat{w}}_i)  (\d{\hat{w}}_j)^* \rangle
\quad\text{and}\quad
[\Swv]_{i} = S_{w_i,v_x} = \langle (\d{\hat{w}}_i) (\d{\hat{v}}_x)^* \rangle,
\label{SwwDefinitionEqn}
\end{equation}
and $S_{vv} = \langle (\d{\hat{v}}_x) (\d{\hat{v}}_x)^* \rangle$.

The structure factor is easily calculated by linearizing  \eqref{MassTransportEqn} and \eqref{MomentumTransportEqn} and transforming into Fourier space,\footnote{The double curl operator is applied to the Fourier transform of \eqref{MomentumTransportEqn} to eliminate the pressure term using the incompressibility constraint~\cite{FluctHydroNonEq_Book}.}
\begin{equation}
\partial_t \V{\hat{\mathcal{U}}} = \V{\mathcal{M}} \V{\hat{\mathcal{U}}} + \V{\mathcal{N}} \V{\hat{\V{\mathcal{Z}}}},
\label{LinearizedTransformedEquations}
\end{equation}
where $\V{\hat{\mathcal{U}}} = (\delta \hat{w}_1,\ldots,\delta \hat{w}_\Nsp, \delta \hat{v}_x)^T$.
This stochastic ODE describes an Ornstein-Uhlenbeck process, so the structure factor is the solution of the linear system~\cite{GardinerBook}
\begin{equation}
\V{\mathcal{M}} \V{S}+\V{S} \V{\mathcal{M}}^* = 
-\V{\mathcal{N}}\V{\mathcal{N}}^*.
\label{OrnsteinUhlenbeckEquation}
\end{equation}
The right hand side is a diagonal matrix with elements,
\begin{equation}
[\V{\mathcal{N}}\V{\mathcal{N}}^*]_{ii} = \frac{2}{\rho}
\left\{\begin{array}{cc} 
k^2 D^0_i m_i \bar{w}_i & i \leq \Nsp \\
k_\perp^2 \nu k_B T & i = \Nsp + 1
\end{array}\right.,
\end{equation}
where $k_\perp^2 = k^2 - k_x^2 = k^2 \sin^2 \theta$, and $\theta$ is the angle between $\V{k}$ and the $x$ axis.

At thermodynamic equilibrium,
\begin{equation}
\V{\mathcal{M}}^{\text{eq}} = 
\left(\begin{array}{c |c}
\V{\mathcal{M}}^{\text{eq}}_{\boldsymbol{w}\boldsymbol{w}} & \boldsymbol{0} \\
\hline
\boldsymbol{0} & -\nu k^2
\end{array}\right),
\end{equation}
where 
\begin{equation}
[\V{\mathcal{M}}^{\text{eq}}_{\boldsymbol{w}\boldsymbol{w}}]_{ij} =
- D^0_i\left(k^2 \delta_{ij}+\frac{z_j}{z_i}\frac{I_i}{\lambda^2 }\right).
\end{equation}
Here the Debye length $\lambda$ is
\begin{equation}
\lambda = \sqrt{\frac{\epsilon k_B T}{\mathcal{I}}},
\qquad\mathrm{where}\qquad
\mathcal{I}=\rho \sum_i m_i w_i z_i^2
\end{equation}
is the ionic strength and $I_i={m_i w_i z_i^2}/\left(\sum_j m_j w_j z_j^2\right)$ is the relative ionic strength.
This can easily be derived from the Fourier transform of the PNP equations. In particular, from \eqref{PoissonEquation} and the condition of local electroneutrality, the fluctuations in the electric field can be expressed in terms of species fluctuations,
\begin{equation}
\delta\hat{\V{E}} = - \iota \V{k} \d{\phi} = - \frac{\iota \V{k}}{\epsilon k^2} \delta \hat{q}
= - \rho\frac{\iota \V{k}}{\epsilon k^2}~\sum_i z_i \delta \hat{w}_i,
\label{EtoW_Equation}
\end{equation}
where $\iota = \sqrt{-1}$.

Solving \eqref{OrnsteinUhlenbeckEquation} at thermodynamic equilibrium gives $\Swveq = 0$ and $S^{\text{eq}}_{vv} = \sin^2(\theta) k_B T/\rho$.  In the case where the solutes are neutral ($V_i = 0$ for all species), which we denote by superscript ``n'', the matrix $\SwweqO$ is diagonal with $S^{\text{(eq,n)}}_{w_i,w_i} = m_i \bar{w}_i/\rho$ independent of $k$. For a mixture involving ionic species,~\cite{LowMachElectrolytes}
\begin{equation}
\Swweq = \SwweqO
-\frac{1}{1+k^2 \lambda^2}~\V{\Pi}
\qquad\mathrm{where}\qquad
\Pi_{i,j} = \frac{\lambda^2}{\epsilon k_B T} \left( m_i z_i \bar{w}_i \right) \left( m_j z_j \bar{w}_j \right).
\label{EquilibriumStructureFactorEquation}
\end{equation}

\subsection{Renormalization of chemical potentials}
\label{sec:RenormChemPot}

It is well-known that the colligative properties (e.g., vapor pressure) of electrolyte solutions depend on their ionic strength, i.e., that the chemical potential of the ions are different from those in a dilute mixture of neutral species. 
Specifically, ionic interactions contribute to the Gibbs free energy and this leads to a correction for the activity. 

The average increase in the electrostatic energy is $\Delta G = \frac{1}{2} \langle \delta q \delta \phi \rangle$. Using (\ref{PlancherelEquation},\ref{EtoW_Equation}) we obtain
\begin{equation}
\Delta G = \frac{\rho^2}{2 \epsilon (2 \pi)^3}  
\int \frac {\V{z}^T (\Swweq - \SwweqO)\, \V{z}}{k^2} ~d \V{k},
\label{eq:DeltaG}
\end{equation}
where we have subtracted $\SwweqO$ to avoid an ill-defined integral that is actually zero due to the overall electroneutrality.
From \eqref{EquilibriumStructureFactorEquation}, the renormalization of the free energy due to fluctuations is
\begin{equation}
\Delta G = -\frac{k_B T} {8\pi \lambda^3}  = - \frac{\mathcal{I}}{8 \pi \epsilon \lambda}.
\label{eq:renorm_G}
\end{equation}
As shown in \cite{robinson2012electrolyte}, this result leads directly to the limiting law of Debye and H\"uckel for point ions.
\modified{
The integration in \eqref{eq:DeltaG} is over all wavenumber; however, FHD is a mesoscopic theory so it does not apply below molecular scales. Introducing an upper bound $k_{\max} \sim \pi/a$, where $a$ is an effective ion radius, reduces the correction $\Delta G$ by a fraction $\sim a/\lambda$ for $a \ll \lambda$, in agreement with the Debye--H\"uckel limiting law for finite-size ions.
}

\section{Fluctuations and Transport}
\label{RenormTransportSection}

Non-equilibrium systems are driven by thermodynamics forces, such as a gradient of concentration or an applied electric field. Transport coefficients such as diffusivity and conductivity are obtained from the linear response, namely the fluxes resulting from weak thermodynamic forces. The linear response is modified due to correlations in the hydrodynamic fluctuations,
\begin{eqnarray}
\bar{\V{F}}_i = \langle \V{F}_i (\V{w},\V{v}) \rangle 
&=& \V{F}_i (\langle \V{w} \rangle, \langle \V{v} \rangle)
+ D_i^0 \frac{eV_i}{k_B T} \langle \delta w_i \delta \V{E} \rangle
+ \langle \delta \V{v} \delta w_i \rangle \nonumber \\
&\equiv& \bar{\V{F}}_i^0 + \bar{\V{F}}_i^\mathrm{relx} + \bar{\V{F}}_i^\mathrm{adv} 
\label{FluctFluxEquation}
\end{eqnarray}
to quadratic order in the fluctuations.
The term $\bar{\V{F}}_i^\mathrm{relx}$ is the relaxation correction and $\bar{\V{F}}_i^\mathrm{adv}$ the advection correction. The term ``relaxation'' refers to the average force experienced by an ion from its asymmetric ionic cloud relaxing due to thermal fluctuations~\cite{robinson2012electrolyte}. 

In what follows we obtain expressions for $\langle \delta w_i \delta \V{E} \rangle$ and $\langle \delta \V{v} \delta w_i \rangle$ from the structure factor and show that
$\bar{\V{F}}_i$ can be written as \eqref{MassFluxEqn} with renormalized diffusion coefficients that depend on the ionic strength. As we shall see, fluctuating hydrodynamics yields the same relaxation and electrophoretic corrections as those obtained by Onsager and co-workers~\cite{OnsagerFuoss1932,OnsagerKim1957},
plus an additional advection enhancement that is given by a Stokes-Einstein formula \cite{DiffusionJSTAT} and is independent of the valences.

For this analysis it is useful to write the matrix $\V{\mathcal{M}}$ as
\begin{equation}
\V{\mathcal{M}} = \V{\mathcal{M}}^{\text{eq}} + \V{\mathcal{M}}' + O(\mathcal{X}^2),
\end{equation}
where $\V{\mathcal{X}}$ is the applied thermodynamic force. In this expansion $\V{\mathcal{M}}^{\text{eq}}$ is $O(\mathcal{X}^0)$ and
$\V{\mathcal{M}}'$ is $O(\mathcal{X}^1)$.
Similarly, we can write the structure factor as $\V{S} = \V{S}^{\text{eq}} + \V{S}' + O(\mathcal{X}^2)$.
The noise covariance matrix $\V{\mathcal{N} \mathcal{N}}^*$ is unchanged,\footnote{This is the so-called local equilibrium assumption, which is valid when the applied gradients are not too large.} so expanding \eqref{OrnsteinUhlenbeckEquation} in powers of $\mathcal{X}$ gives 
the correction to the structure factors to linear order in $\V{\mathcal{X}}$ as the solution of the linear system
\begin{equation}
\V{\mathcal{M}}^{\text{eq}} \V{S}' + \V{S}' (\V{\mathcal{M}}^{\text{eq}})^* = -\V{\mathcal{M}}' \V{S}^{\text{eq}} - \V{S}^{\text{eq}} (\V{\mathcal{M}}')^*.
\label{neq_OU}
\end{equation}


\subsection{Renormalization of diffusion}
\label{RenormDiffusionSection}

For neutral species, diffusion can be analyzed by imposing a concentration gradient separately for each species and formulating the linearized response. For charged species, however, the concentration gradient of a given species must be balanced by the other concentration gradients in order to preserve electroneutrality in the mean, $\sum_i z_i \grad \bar{w}_i = 0$; as mentioned, we assume all concentration gradients are in the $x$-direction.

For an imposed concentration gradient $\V{\mathcal{X}} \equiv \nabla_x \bar{\V{w}}$ in the absence of an external electric field ($\langle \V{E} \rangle = 0$), the relaxation of composition fluctuations follows the linearized equations
\begin{equation}
\partial_t \delta w_i =
D_i^0 \left(\nabla^2 \delta w_i - \frac{I_i}{\lambda^2 z_i}\sum_{j} z_j \delta w_j \right)
- \frac{D_i^0 m_i z_i}{k_B T} \d{E}_x \nabla_x \bar{w}_i 
 - \delta v_x \nabla_x \bar{w}_i. 
\end{equation}
The linearized momentum equation is the same as in equilibrium.
Using \eqref{EtoW_Equation} we obtain the linear correction to the relaxation matrix,
\begin{equation}
\V{\mathcal{M}}'=
\left(\begin{array}{c | c}
\iota \frac{\cos\theta}{k} \frac{\rho}{\epsilon k_B T} \V{\pi}\V{z}^T
& -\nabla_x \bar{\V{w}} \\
\hline 
\V{0} & 0
\end{array}
\right),
\label{M_GradW_Eqn}
\end{equation}
where the column vector $\V{\pi}$ has elements $\pi_i = D_i^0 m_i z_i \nabla_x \bar{w}_i$.

\subsubsection{Advective correction}
\label{sec:ElphorDiff}

The advective correction $\bar{\V{F}}_i^\mathrm{adv}$ to the fluxes due to nonzero correlation $\langle \d{\V{v}} \d{w_i} \rangle$ in \eqref{FluctFluxEquation} can be calculated rather easily since $\SwvP$ solves
\begin{equation}
\V{\mathcal{M}}^{\mathrm{eq}}_{\boldsymbol{w}\boldsymbol{w}} \SwvP 
- \nu k^2 \SwvP = \frac{k_B T}{\rho} \sin^2 \theta ~\nabla_x \bar{\V{w}}.
\end{equation}
Using the constraint $\V{z}^T \nabla_x \bar{\V{w}}=0$ gives 
\begin{equation}
\SwvP = -\frac{k_B T \sin^2 \theta}{k^2 \rho} \text{Diag}(D_i^0+\nu)^{-1} ~\nabla_x \bar{\V{w}}.
\end{equation}
Integrating over $k$ and using \eqref{PlancherelEquation} then yields
\begin{equation}
\bar{\V{F}}_i^\mathrm{adv} = \langle \delta \V{v} \delta w_i \rangle 
= - \frac{k_B T}{3 \pi \rho (D_i^0+\nu) a_i} \grad \bar{w}_i,
\label{diffusionRenormalization}
\end{equation}
where we have introduced a molecular length $a_i$ to set an upper bound of $\pi/a_i$ for the wavenumber in order for the integral \eqref{PlancherelEquation} to converge. One can interpret $a_i$ as the molecular hydrodynamic diameter that enters in the Stokes-Einstein formula \cite{DiffusionJSTAT}.

The advective contribution to the fluxes~\eqref{diffusionRenormalization} can be absorbed into the PNP equations by redefining or renormalizing the diffusion coefficients from their bare values $D^0_i$ to
\begin{equation}
    D_i = D^0_i + \frac{k_B T}{3 \pi \mu a_i},
    \label{RenormalizedD_i}
\end{equation}

\subsubsection{Relaxation correction}
\label{sec:RelaxDiff}

The relaxation correction $\bar{\V{F}}_i^\mathrm{relx}$ due to the nonzero correlation $\langle \delta w_i \delta \V{E} \rangle$ in \eqref{FluctFluxEquation} can be computed in principle by solving for $\SwwP$ the linear system
\begin{equation}
\V{\mathcal{M}}^{\mathrm{eq}}_{\boldsymbol{w}\boldsymbol{w}} \SwwP
+ \SwwP (\V{\mathcal{M}}^{\mathrm{eq}}_{\boldsymbol{w}\boldsymbol{w}})^T
= -\iota~\frac{\cos\theta}{k}
\frac{\rho}{\epsilon k_B T}
(\V{\pi} \V{z}^T  \Swweq - \Swweq \V{z} \V{\pi}^T).
\label{general_eqs}
\end{equation}
This can be simplified further to the system
\begin{equation}
    \M{D}\M{\Omega}\SwwP + \SwwP \M{\Omega}^T \M{D}  = 
    \iota~\frac{k \lambda^2 \cos\theta}{\rho \left( 1+k^2\lambda^2 \right)}
    \left( \V{\pi}\V{\kappa}^T-\V{\kappa}\V{\pi}^T \right),
    \label{general_eqs_simplified}
\end{equation}
where $\M{D}=\text{Diag}(D^0_i)$, the column vector $\V{\kappa}$ has elements $\kappa_i = m_i w_i z_i $, and the matrix $\M{\Omega}= k^2 \lambda^2 \left(\V{z}^T \V{\kappa}\right) \M{I} + \V{\kappa}\V{z}^T$. The solution $\SwwP$ is a purely imaginary anti-symmetric matrix with zeros on the diagonal.

Given a solution to~\eqref{general_eqs}, we can use (\ref{SwwDefinitionEqn},\ref{EtoW_Equation}) to obtain
\begin{equation}
\bar{\V{F}}_i^\mathrm{relx} = 
 D_i^0 \frac{eV_i}{k_B T} \langle \delta w_i \delta \V{E} \rangle
= 
 - D_i^0 \frac{eV_i}{\epsilon k_B T} \frac{\rho}{8 \pi^3} \sum_{j} z_j \int d\V{k}~
\frac{\iota \V{k}}{k^2} S^\prime_{w_i,w_j}.
\label{average_wE}
\end{equation}
\modified{Since all computations are linear, the final result can be written as a correction to Fick's law,
$\bar{\V{F}}_{x}^\mathrm{relx} = - \M{D}^\mathrm{relx} \nabla_x \bar{\V{w}}$, 
where, in general, $\M{D}^\mathrm{relx}$ is not diagonal and includes cross-diffusion terms. As done earlier with \eqref{eq:DeltaG}, introducing an upper bound $k_{\max} \sim \pi/a$ in \eqref{average_wE} reduces the relaxation correction by a fraction $\sim a/\lambda$, in agreement with Onsager's calculations for finite-size ions.}
One can also express the results in terms of corrections to the binary Maxwell-Stefan diffusion coefficients \modified{(see Section~\ref{sec:ternary} and the Appendix)}~\cite{ElectrolytesMS_Review,PnasElectrolyte}.

In general, it is difficult to solve \eqref{general_eqs} in closed form; an explicit but lengthy formulation for the relaxation correction to diffusion is given by Onsager and Kim~\cite{OnsagerKim1957} in terms of solutions to eigenvalue problems.\footnote{Note that in our calculation only a linear system needs to be solved and integrals performed, without actually computing eigenvalues.} In ~\ref{sec:binary} we give explicit results for a binary electrolyte, and in Section~\ref{sec:ternary} for a symmetric ternary electrolyte mixture.

\subsection{Renormalization of conductivity}
\label{RenormConductivitySection}

By Ohm's law, the electrical conductivity $\Lambda_i$ for species $i$ is given by
$z_i \bar{\V{F}_i} = \Lambda_i \V{E}_{\text{ext}}$,
where $\V{E}_{\text{ext}}$ is the applied electric field.
In \cite{PnasElectrolyte}, we derived the renormalization of the conductivity of a 1:1 electrolyte solution with ions of equal mobility. To generalize that result we follow the same procedure as for the renormalization of the diffusion coefficients, except that instead of imposing concentration gradients we apply an external electric field $\V{\mathcal{X}} \equiv \V{E}_{\text{ext}}=E_\text{ext} \V{e}_x$.
From the linearized PNP equations in the presence of an applied field one can easily obtain
\begin{equation}
\V{\mathcal{M}}' =
\left(\begin{array}{c | c}
-\iota \frac{k \cos\theta}{k_B T} \V{\alpha} & \V{0}\\ 
\hline
E_{\text{ext}}  \sin^2(\theta) \V{z}^T & 0
\end{array}
\right),
\label{M_Efield_Eqn}
\end{equation}
where $\V{\alpha} = \text{Diag}\left( D_i^0 m_i z_i E_{\text{ext}}\right)$.

\subsubsection{Advective correction}

As in the derivation in subsection \ref{sec:ElphorDiff}, the advective correction to the fluxes is computed by solving for $\SwvP$ the linear system
\begin{equation}
\V{\mathcal{M}}^{\mathrm{eq}}_{\boldsymbol{w}\boldsymbol{w}} \SwvP - \nu k^2 \SwvP =  
- \frac{\lambda^2 k^2 \sin^2 \theta}{1+\lambda^2 k^2} ~\SwweqO \V{z} E_{\text{ext}},
\end{equation}
to obtain
\begin{equation}
S^\prime_{w_i,v} = \frac{\lambda^2 \sin^2 \theta  }{1+\lambda^2 k^2}~ \frac{m_i \bar{w}_i z_i}{\rho (D_i^0 + \nu)} \, E_{\text{ext}}.
\end{equation}
This gives via \eqref{PlancherelEquation} the flux correction
\begin{equation}
\bar{\V{F}}_i^\mathrm{adv} = \langle \delta \V{v} \delta w_i \rangle 
\approx \left( \frac{1}{3\pi a_i} - \frac{1}{6 \pi \lambda} \right) \frac{m_i \bar{w}_i z_i}{\mu} ~\V{E}_{\text{ext}},
\label{AdvectionConductivityGeneralEqn}
\end{equation}
\modified{for Schmidt number $\text{Sc} \gg 1$} and $\lambda \gg a$, as suitable for dilute solutions in a liquid.

The advection contribution to the conductivity coming from \eqref{AdvectionConductivityGeneralEqn} has two terms. The first contribution involves the molecular cutoff $a$ and is consistent with the renormalization of the diffusion coefficient in \eqref{RenormalizedD_i}. 
The second contribution involves the Debye length and is precisely the electrophoretic term obtained by Onsager and Fuoss~\cite{OnsagerFuoss1932};  it leads to strong cross-species corrections to the PNP equations of order square root in the ionic strength.

\subsubsection{Relaxation correction}

For the relaxation contribution we need to solve for $\SwwP$ the system
\begin{equation}
\V{\mathcal{M}}^{\mathrm{eq}}_{\boldsymbol{w}\boldsymbol{w}} \SwwP 
+ \SwwP (\V{\mathcal{M}}^{\mathrm{eq}}_{\boldsymbol{w}\boldsymbol{w}})^* =
-\iota ~ \frac{k \cos(\theta)}{k_B T(1+k^2 \lambda^2)}\left(\V{\alpha}~{\V{\Pi}} - {\V{\Pi}}~\V{\alpha} \right).
\label{general_eqs_E}
\end{equation}
This can be simplified further to the system
\begin{equation}
    \M{D}\M{\Omega}\SwwP + \SwwP \M{\Omega}^T \M{D}  = 
    \iota~\frac{k \lambda^2 \cos\theta}{\rho k_B T \left( 1+k^2\lambda^2 \right)}
    \left( \V{\omega}\V{\kappa}^T-\V{\kappa}\V{\omega}^T \right) E_{\text{ext}},
    \label{general_eqs_E_simplified}
\end{equation}
where we used the same notation as in \eqref{general_eqs_simplified}, and the column vector $\V{\omega}$ has elements $\omega_i = D^0_i m_i^2 z_i^2 w_i$.  The solution $\SwwP$ is a purely imaginary anti-symmetric matrix with zeros on the diagonal.

After solving for $\SwwP$, the flux correction can be obtained by performing an integral over $\V{k}$. We give explicit results in~\ref{sec:binary} for a general binary case, and in Section~\ref{sec:ternary} for a symmetric ternary case.

\section{Symmetric Ternary Ion Mixture}
\label{sec:ternary}

We now consider a ternary system with one cation and two anions (valences $V_1 = +1$, $V_2 = V_3 = -1$) in a solvent. To simplify the analysis we take the ions to have equal masses $m_i=m$ (and thus equal charges per mass $z=e/m$) and ``bare'' diffusivity $D^0_i=D^0$. By the electroneutrality condition the average composition of the mixture can be written as $\bar{\V{w}} = w_0 (1, f, 1-f)^T$, where $f(\V{r})$ is the relative fraction between the two anions.


For this symmetric ternary mixture, instead of the concentrations $(w_1, w_2, w_3)$, we introduce as variables:
the total mass fraction of solutes $n=w_1 + w_2 + w_3$, the mass fraction of net charge $c=w_1 - (w_2 + w_3)$, and the difference in mass fractions between the anions $s=w_2 - w_3$.
These have average values $\langle n \rangle = 2 w_0$, $\langle c \rangle = 0$, and $\langle s \rangle = w_0 (2 f - 1)$.
The corresponding hydrodynamic fluxes are
\begin{equation}
\left(
\begin{array}{c}
\V{F}_{n} \\ \V{F}_{c} \\ \V{F}_{s}
\end{array}
\right)
=
-D^0
\left(
\begin{array}{c}
\grad n \\ \grad c \\ \grad s
\end{array}
\right)
+
\frac{D^0 m z}{k_B T} \V{E}
\left(
\begin{array}{c}
c \\ n \\ - s
\end{array}
\right)
+
\V{v}
\left(
\begin{array}{c}
n\\
c\\
s
\end{array}
\right).
\end{equation}
In the basis $(n,c,s,v_x)$ the equilibrium matrix $\V{\mathcal{M}}^\mathrm{eq}$ is
\begin{equation}
\V{\mathcal{M}}^{\text{eq}} = 
\left(
\begin{array}{c c c c}
-D^0 k^2 & 0 & 0 & 0\\
0 & -D^0 k^2 - D^0\lambda^{-2} & 0 & 0 \\
0 & -\frac{1}{2}D^0(1-2f)\lambda^{-2} & -Dk^2 & 0 \\
0 & 0 & 0 & -\nu k^2
\end{array}
\right),
\end{equation}
and the noise covariance matrix is
\begin{equation}
\V{\mathcal{N N}^*} = \frac{2 k^2 D^0 m w_0}{\rho}
\left(
\begin{array}{c c c c}
2  & 0 &  (2f-1) & 0 \\
0 & 2  &   (1-2f) & 0 \\
 (2f -1) &   (1-2f) &  1 & 0 \\
0 & 0 & 0 &  \frac{\sin^2 ( \theta ) \nu k_B T}{D^0 m w_0}
\end{array}
\right).
\end{equation}

\subsection{Renormalization of diffusion coefficients}


We first set the applied electric field to zero and write the gradients of concentrations
in terms of the independent gradients of saltiness $\grad \bar{w}_0$ and label (or color) of the anions $\grad f$, 
\begin{equation}
\nabla_x \bar{\V{w}} = \left( \nabla_x \bar{w}_0,\,  f \left(\nabla_x \bar{w}_0\right) + \bar{w}_0 (\nabla_x f),\, (1-f) \left(\nabla_x \bar{w}_0\right) - \bar{w}_0 (\nabla_x f) \right).
\label{grad_w_bar}
\end{equation}
In the basis $(n,c,s,v_x)$,
\begin{equation}
\V{\mathcal{M}}' =
\left(
\begin{array}{c c c c}
0 & 0 & 0 & -2 \left(\nabla_x \bar{w}_0\right) \\
0 & \iota \frac{\cos ( \theta ) D^0 }{k \bar{w}_0 \lambda^2} \left(\nabla_x \bar{w}_0\right) & 0 & 0  \\
0 & \iota \frac{\cos ( \theta ) D^0 }{2 k \bar{w}_0 \lambda^2} g_{12} & 0 & g_{12} \\
0 & 0 & 0 & 0
\end{array}
\right),
\end{equation}
where $g_{12}= (1-2 f) \left(\nabla_x \bar{w}_0\right) - 2 \bar{w}_0 \left(\nabla_x f\right)$.

The relaxation contributions to the fluxes are obtained from \eqref{average_wE},
\begin{equation}
\bar{\V{F}}_n^{\text{relx}} = \bar{\V{F}}_c^{\text{relx}} = \V{0}, \qquad
\bar{\V{F}}_s^{\text{relx}} = 
\frac{D^0 m (2-\sqrt{2}) }{24 \pi \rho \lambda^3 } \grad f.
\label{relaxation_diffusion_ternary}
\end{equation}
The advective contributions are obtained from \eqref{diffusionRenormalization}
and are simply found to be consistent with the PNP equations after a renormalization of the diffusion coefficients according to \eqref{RenormalizedD_i}.
In particular, there is no electrophoretic contribution to $\bar{\V{F}}^\mathrm{adv}$ when the thermodynamic forces are due to concentration gradients.

\subsection{Renormalization of conductivity}

Next we consider the case where there is an applied electric field with $\grad n = \grad c =\grad s = \V{0}$. In this case,
\begin{equation}
\V{\mathcal{M}}' = \left(
\begin{array}{c c c c}
0 & -\iota\frac{D^0 m z }{k_B T} k \cos(\theta) & 0 & 0 \\
-\iota\frac{D^0 m z }{k_B T} k \cos(\theta) & 0 & 0 & 0 \\
0 & 0 & \iota \frac{D^0 m z}{k_B T} k \cos(\theta) & 0 \\
0 & z \sin^2 ( \theta ) & 0 & 0
\end{array}
\right)\V{E}_\mathrm{ext}.
\end{equation}
After solving for the structure factor deviation $\V{S}'$, 
the hydrodynamic fluxes of quantities $n$, $c$, and $s$ can be evaluated.
The contributions from the relaxation term are
\begin{equation}
\left(\begin{array}{c}
\bar{\V{F}}^{\text{relx}}_n \\
\bar{\V{F}}^{\text{relx}}_c \\
\bar{\V{F}}^{\text{relx}}_s
\end{array}\right)
=  \frac{D^0 m^2 z (2-\sqrt{2})}{48 \pi k_B T \rho \lambda^3 }
\left(\begin{array}{c}
0 \\
-2 \\
2 f-1
\end{array}\right)
\V{E}_\mathrm{ext},
\label{relaxation_conductivity_ternary}
\end{equation}
and \modified{for large Schmidt number} the contributions from the advection term are
\begin{equation}
\left(\begin{array}{c}
\bar{\V{F}}^{\text{adv}}_n \\
\bar{\V{F}}^{\text{adv}}_c \\
\bar{\V{F}}^{\text{adv}}_s
\end{array}\right) \approx
\left( \frac{1}{3\pi a} - \frac{1}{6 \pi \lambda} \right) \frac{m w_0 z}{\mu} 
\left(\begin{array}{c}
0 \\
2 \\
1-2f
\end{array}\right)~\V{E}_{\text{ext}}.
\label{advective_conductivity_ternary}
\end{equation}
The first part of this correction ($\sim a^{-1}$) can be absorbed into the PNP equations with a renormalization of the diffusion coefficients according to \eqref{RenormalizedD_i}, while the second part ($\sim \lambda^{-1}$) is an electrophoretic correction that is consistent with the calculations of Onsager and coworkers~\cite{OnsagerFuoss1932,OnsagerKim1957}.

\subsection{Renormalization of Maxwell-Stefan coefficients}

From (\ref{relaxation_diffusion_ternary},\ref{relaxation_conductivity_ternary},\ref{advective_conductivity_ternary}), one can obtain the complete \emph{non-diagonal} Fickian diffusion matrix for the ternary electrolyte.
In the Fickian formulation the gradient of chemical potentials is represented by the vector $\V{g}$ with $g_i = \grad \bar{w}_i + (e V_i \bar{w}_i) (\grad \bar{\phi})/(k_B T)$ (c.f. \eqref{MassFluxEqn}).
Specifically, we can write \eqref{FluctFluxEquation} in the form
\begin{equation}
    \bar{\V{F}} = - \V{D}_{\text{Fick}} \,\V{g} = 
    - \left( D\,\M{I} + \V{D}^{\text{adv}} + \V{D}^{\text{relx}} \right) \,\V{g}
    \label{renormalized_Fick_matrix}
\end{equation}
where the renormalized diffusion coefficient is $D = D^0+k_B T /(3 \pi \mu a)$.
 The advective correction to the Fickian diffusion matrix is
\begin{equation}
\V{D}^{\text{adv}} = 
-\frac{k_B T}{12 \pi \lambda \mu}
\left(
\begin{array}{c c c}
 1 & -1 & -1\\
- f & f & f\\
(-1 + f) & 1-f & 1-f
\end{array}
\right),
\end{equation}
while the relaxation correction is
\begin{equation}
\V{D}^{\text{relx}} = 
-\frac{m D^0 (2 -\sqrt{2})}{96 \pi \lambda^3 \rho w_0}
\left(
\begin{array}{c c c}
1 & -1 & -1 \\
- f &  2-f & - f\\
-1+f & -1+f & 1+f
\end{array}
\right).
\end{equation}

One can convert the Fickian coefficients to binary Maxwell-Stefan coefficients $\text{\DJ}_{ij}$, which are preferred to the Fickian diffusion coefficients \cite{ElectrolytesMS_Review}. To leading order in the ionic strength,
\begin{equation}
\text{\DJ}_{ij} = -12 \pi V_i V_j D^2 \left[ \frac{k_B T}{\mu} 
- V_i V_j \frac{m D^0 (2 -\sqrt{2})}{8 \rho \bar{w}_0 \lambda^2}  \right]^{-1} \frac{m_s}{m} \lambda \bar{w}_0,
\label{negative_MS}
\end{equation}
where $m_s$ is the mass of the solvent molecules and the valencies are plus or minus one.
Equation~\eqref{negative_MS} can be considered the final result of our calculation since once the full MS diffusion matrix is known one can compute all of the fluxes given the thermodynamic driving forces.

\modified{It is important to note that both the electrophoretic and relaxation corrections to the transport coefficients are proportional to the inverse square root of the ionic strength. They are both positive for counter-ions but have opposite signs for co-ions. Furthermore, the co-ion MS coefficient in a symmetric ternary mixture can be negative (without violating any laws of physics), unlike the MS coefficients between uncharged species. This result is corroborated by experimental measurements showing that the Maxwell-Stefan coefficient between co-ions can be negative \cite{ElectrolytesMS_Review,NegativeMaxStefanDiff1,NegativeMaxStefanDiff2}. Finally, the fact that the MS coefficients both depend on concentration very strongly and can be negative makes them less attractive for electrolyte mixtures; however, we are not aware of any better alternatives.}


\section{Concluding Remarks}
\label{ConclusionSection}

Fluctuating hydrodynamics is a powerful modeling tool at mesoscopic scales, as demonstrated here by the calculation of the thermodynamic and transport corrections for electrolytes originally derived by Debye, H\"uckel, Onsager, and co-workers. Our straightforward calculations showed that the (fluctuating) PNP equations need to be corrected to order square root in the ionic strength, and are thus valid only for very dilute solutions. The FHD formulation reveals the physics behind these corrections, such as the electrophoretic correction in conductivity arising from a correlation between velocity and concentration fluctuations. Yet fluctuating hydrodynamics has many applications beyond this elegant and insightful formulation of classic results. One of the strengths of FHD is being able to model complex non-ideal multi-species mixtures including contributions due to mean fluid flow, temperature gradients, boundary conditions (e.g., see~\cite{DDFT_Electrolytes_Dean}), etc.

The present analysis was taken only to linear order but the extension to higher order is possible (e.g., corrections for strong fields are predicted for the relaxation term~\cite{DDFT_Electrolytes_Dean}).
Using ``one-loop'' renormalization theory, in this work we were only able to compute the leading-order corrections $\sim \sqrt{\mathcal{I}}$ in the ionic strength $\mathcal{I}$.
Theoretical approaches have also been developed to go beyond dilute solutions~\cite{ICEO_Concentrated_Bazant}. Logarithmic corrections to transport coefficients have been computed by Chen and Onsager~\cite{Electrolytes_OnsagerChen}, but to our knowledge the exact formula for the coefficients is not yet agreed upon~\cite{ElectrolytesDynamics_Review}.  It remains to be seen if a higher-order (analytical or numerical) perturbation analysis of the FHD equations can produce higher-order logarithmic corrections $\sim \mathcal{I} \ln \mathcal{I}$~\cite{Electrolytes_OnsagerChen}. The next order terms are affected by molecular details such as ion pair formation~\cite{pikal1971ion,ElectrolytesDynamics_Review} and are thus likely beyond the reach of hydrodynamic theories.

It is important to point out that at higher concentrations one must use a complete multicomponent transport model including nonideality of the solution and the flux of the solvent that comes from the conservation of mass~\cite{ElectrolytesMS_Review,ICEO_Concentrated_Bazant,NernstPlanckModel}. In our work~\cite{LowMachElectrolytes} we have summarized the complete mass and momentum transport equations consistent with nonequilibrium thermodynamics, without the need to single out a solvent species or assume a dilute solution. Ion crowding can be modelled with additional (e.g. fourth order) terms not included in traditional models~\cite{IonCrowdingBSK_Bazant}.
Here we treated strong electrolyte solutions but the extension to weak electrolytes, as well as general electrochemistry, is straight-forward. Another important extension of this work is to consider the AC conductivity of electrolytes as a function of frequency~\cite{Electrolytes_Falkenhagen}; this is in principle a straighforward but tedious extension of the approach.

\modified{We should also mention some limitations of our calculations.} In the analytical perturbative approach followed here, all corrections to the linearized fluctuating PNP equations appear additively, not multiplicatively as they should.
For example, the $\V{g}$ appearing in~\eqref{renormalized_Fick_matrix} should include contributions from~\eqref{eq:renorm_G}. 
Similarly, we wrote the relaxation corrections in terms of bare diffusion coefficients $D^0$, but in reality the bare coefficients are simultaneously renormalized by the advective correction.
Multiplicative effects like these can be important and are easily computed by nonlinear computational fluctuating hydrodynamics (e.g., see ~\cite{DiffusionJSTAT}).

Numerical methods for solving the FHD equations are well-established and these are especially useful for including the effects of boundary conditions (e.g., conductivity corrections are predicted for the relaxation term in a confined binary mixture~\cite{DDFT_Electrolytes_Dean}) and nonlinearities. Care is needed, however, to avoid double-counting the contribution of the fluctuations to transport. In computational formulations the renormalization depends on the coarse-graining scale (e.g., grid size) and this must be carefully evaluated by stochastic numerical analysis. With this caveat, fluctuating hydrodynamics promises to be a powerful mesoscopic formulation for electrolytes.

\section*{Acknowledgements}

We thank Martin Bazant for invaluable advice on how to place our work in the broader context of electrolyte modeling.
This work was supported by the
U.S.~Department of Energy, Office of Science,
Office of Advanced Scientific Computing Research,
Applied Mathematics Program under award Award DE-SC0008271 and contract DE-AC02-05CH11231.
A. Donev was supported in part by the Division of Chemical, Bioengineering, Environmental and Transport Systems of the National Science Foundation under award CBET-1804940.

\appendix

\section{Binary Electrolytes}
\label{sec:binary}

In this appendix, we generalize the binary electrolyte theory presented in \cite{PnasElectrolyte} by letting the two ionic species have different physical properties, in particular, unequal diffusion coefficients. We still assume equal valences, $V_1=V_2=V$, because it simplifies the expressions greatly while retaining the interesting features.
The general expressions for the advective contributions due to the $\langle \delta w_i \delta \V{v} \rangle$ term were derived in Sections~\ref{RenormDiffusionSection} and \ref{RenormConductivitySection} so here we focus on the relaxation contribution from the $\langle \delta w_i \delta \V{E} \rangle$ term. 

We first consider an applied concentration gradient $\grad_x \bar{\V{w}} = ( 1, m_2/m_1) \nabla_x \bar{w}_1$. The calculation of $\V{S}'$ is tedious but straight-forward; the diagonal terms are zero, and the off-diagonal terms are
\begin{equation}
S_{12}' = -\iota \frac{m_2 (D_1^0-D_2^0) }{\rho(D_1^0+D_2^0)} 
\frac{k \lambda^2  \cos(\theta)}{(k^2 \lambda^2+1)(2 k^2 \lambda^2 +1)} ~\nabla_x \bar{w}_1.
\label{S12BinaryEquation}
\end{equation}
Using this and (\ref{average_wE}) the relaxation contribution to the flux of species 1 is
\begin{equation}
\bar{\V{F}}_1^\mathrm{relx} = \frac{D_1^0 m_1 z_1}{k_B T} \langle \delta w_1 \delta \V{E} \rangle  = - D_1^\mathrm{relx} \grad w_1,
\end{equation}
where
\begin{equation}
D_1^\mathrm{relx} = \frac{(2-\sqrt{2}) m_1 D_1^0  (D_2^0 -D_1^0)}{48 \pi \rho \bar{w}_1 \lambda^3 (D_1^0 +D_2^0)},
\end{equation}
with a similar expression for species 2. As expected, this contribution is zero when the mobilities are equal.

We now consider an applied external electric field. Performing the calculation from Section~\ref{RenormConductivitySection} we obtain
\begin{equation}
\bar{\V{F}}_i^\mathrm{relx} = \frac{D_i^0 m_i z_i}{k_B T} 
\langle \delta w_i \delta \V{E} \rangle 
= - \frac{(2-\sqrt{2}) D_i^0 m_i^2 z_i}{48 \pi k_B T \rho \lambda^3}  ~\V{E}_\mathrm{ext},
\label{RelaxationContributonBinaryEqn}
\end{equation}
which is in exact agreement with the result obtained by Onsager and Fuoss.


From these results, one can compute the Maxwell-Stefan binary diffusion coefficients. 
To leading order in ionic strength, the corrections to the pairwise diffusion coefficients between the ions and the solvent can be shown to have an electrophoretic correction for unequal ions,
\begin{equation}
\text{\DJ}_{1s} = D_1+\frac{k_B T (D^0_1-D^0_2)}{12 \pi \lambda \mu D^0_2},
\end{equation}
and similarly for $\text{\DJ}_{2s}$. The cross-diffusion coefficient between the two ions has both an electrophoretic and a relaxation correction,
\begin{equation}
\text{\DJ}_{12} = 12 \pi D_1 D_2 \left[ \frac{k_B T}{\mu} 
+ \frac{(2-\sqrt{2}) m_1}{4 \rho w_1 \lambda^2}\frac{D^0_1 D^0_2}{D^0_1 + D^0_2}  \right]^{-1} \frac{m_s}{m_1} \lambda \bar{w}_1
\label{BinaryMSdiffusionCoefficientEquation}
\end{equation}
\modified{so $\text{\DJ}_{12} \propto \mathcal{I}^{-1/2}$.}  This is the generalization of the result for symmetric ions derived in \cite{PnasElectrolyte}, and matches~\eqref{negative_MS} for dynamically-identical counter-ions.
As explained in the Conclusions, the bare diffusion coefficients in these formulas should actually be replaced by their renormalized values throughout once one accounts for nonlinear (multiplicative) effects.

\bigskip\hrule\bigskip


\end{document}